\documentclass[default]{sn-jnl}
\usepackage{graphicx}%
\usepackage{multirow}%
\usepackage{amsmath,amssymb,amsfonts}%
\usepackage{amsthm}%
\usepackage{mathrsfs}%
\usepackage[title]{appendix}%
\usepackage{xcolor}%
\usepackage{textcomp}%
\usepackage{manyfoot}%
\usepackage{booktabs}%
\usepackage{algorithm}%
\usepackage{algorithmicx}%
\usepackage{algpseudocode}%
\usepackage{listings}%
\usepackage{arydshln}
\theoremstyle{thmstyleone}%
\theoremstyle{thmstyletwo}%

\theoremstyle{thmstylethree}%

\begin{document}

\title[Highly efficient coupling of single photons using a pair of nanostructures]{Highly efficient coupling of single photons using a pair of nanostructures}
\author[1]{\fnm{Resmi} \sur{M}}\email{19phph18@uohyd.ac.in}

\author[1]{\fnm{Elaganuru} \sur{Bashaiah}}\email{18phph19@uohyd.ac.in}
\author[1]{\fnm{Shashank} \sur{Suman}}\email{shashanksuman830@gmail.com}
\author*[1]{\fnm{Ramachandrarao} \sur{Yalla}}\email{rrysp@uohyd.ac.in}
\affil[1]{\orgdiv{School of Physics}, \orgname{University of Hyderabad}, \orgaddress \city{Hyderabad}, \postcode{500046}, \state{Telangana}, \country{India}}

\abstract{We numerically report highly efficient coupling of single photons from a single dipole source (SDS) using a pair of nanostructures. The maximum coupling efficiency ($\eta_{p}$) of 56\%, into guided modes of a silica nanotip (SNT), is found when the SNT of radius 0.43 $\mu m$ is placed in the vicinity of a diamond nanotip (DNT) and a diamond nanowire (DNW). Additionally, we found that varying the radius of the DNT/DNW does not significantly affect the $\eta_{p}$-value. Furthermore, we investigate the coupling efficiency ($\eta$) of single photons from the SDS into guided modes of the DNT. The maximum $\eta$-value of 87\% is found when the radially oriented SDS is positioned on the facet of the DNT of radius 0.4 $\mu m$. We found that the $\eta_{p}$-value is enhanced when the DNT is placed in the vicinity of another DNT and the DNW. The present platform may open new possibilities in quantum networks.}

\maketitle

\section{Introduction}\label{Intro}
The single photons emitted by a quantum emitter, which are efficiently coupled into guided modes of a single-mode fiber is a crucial requirement for robust applications in quantum technologies, including quantum computation \cite{kimble2008quantum,o2009photonic,o2007optical}. Various solid-state quantum emitters have been investigated to generate single photons in an efficient way \cite{aharonovich2016solid,kiraz2004quantum}. Various groups worldwide have demonstrated the collection of single photons using a variety of micro/nano-structures. Examples include photonic crystal cavities \cite{badolato2005deterministic}, micropillar cavities \cite{solomon2001single}, high finesse optical cavities in free-space \cite{hood2000atom}, optical nanocapillary fibers \cite{bashachanneling}, and diamond nanobeams  \cite{hausmann2013coupling}. Although these structures are efficient, a fiber-integrated platform is in high demand considering its applications in fiber-based networks. In this regard, silica nanowires (SNW) \cite{wu2013optical} and silica nanotips (SNT) \cite{berneschi2020optical} became promising candidates for the above-said purpose. Recently, diamond based nanostructures like diamond nanowires (DNW), diamond nanotips (DNT) and diamond nanobeams are also emerging platforms in quantum technologies.

A single atom placed in the vicinity of the SNW could achieve the maximum coupling efficiency ($\eta$) of 28\% \cite{le2005spontaneous}. The same has been demonstrated experimentally using a single cesium atom \cite{nayak2008single}. The maximum $\eta$-value of 22\% is experimentally demonstrated by depositing a single quantum dot (QD) on the surface of the SNW \cite{yalla2012efficient}. The maximum $\eta$-value of 38\% has been reported numerically for the SNT and a single dipole source (SDS) in the single-mode regime \cite{chonan2014efficient}. Recently, our group reported the $\eta$-value of 43.5\% for the SNT in the multi-mode regime \cite{resmi2023efficient,resmichanneling}. The SDS is placed on the facet of the SNT, at its centre, and radius of the SNT is varied to determine the $\eta$-value dependence. The three peak $\eta$-values correspond to the SNT radius of 0.20 $\mu m$ (P$_{1}$), 0.43 $\mu m$ (P$_{2}$), and 0.71 $\mu m$ (P$_{3}$), respectively. The corresponding $\eta$-values are 37\%, 43.5\% and 43\%, respectively. The efficient coupling of single photons into the DNW has been experimentally demonstrated \cite{babinec2010diamond}. The maximum $\eta$-value of 61\% is numerically demonstrated for a SDS on the DNT \cite{das2023efficient}.

However, to enhance the $\eta$-value, other nanostructures and hybrid systems have been integrated with SNWs, SNTs, and DNWs. A single QD in a cavity formed by the combination of the SNW and a nanofabricated grating is reported to be efficient  \cite{yalla2014cavity}. The Bragg cavity introduced on the SNW is found to have enhanced $\eta$-value up to 84\% \cite{schell2015highly,takashima2016detailed}. The experimental results have shown the maximum $\eta$-value of 16-37\% for a hybrid structure consisting of the SNW and a DNW with a nitrogen-vacancy (NV) center in it \cite{patel2016efficient}. Although, for this hybrid system, numerical simulations have shown that the $\eta$-value reaches a maximum of $\sim$75\%  \cite{yonezu2017efficient}. The $\eta$-value of 80\% is demonstrated by introducing a narrow air-filled groove into the cavity on the SNW \cite{li2018tailoring}. Single atoms interfaced on photonic crystal structures fabricated on the SNW is found to have the maximum $\eta$-value of 85\% \cite{nayak2019real}. It has been numerically shown that placing the single quantum emitter in a hole etched on the SNW can enhance the $\eta$-value up to $\sim$63\% \cite{wang2021high}. The integration of silicon-vacancy (SiV) centers in nanodiamonds with SNW has been experimentally demonstrated with the $\eta$-value of 4\%. \cite{yalla2022integration}. Numerical simulations have predicted an increase in the $\eta$-value upto 43\% when a gold nanoparticle is in the proximity to the SNT \cite{das2023efficient}. It is also shown that the $\eta$-value gets enhnaced to 64\% when the SDS-DNT system is in the proximity of a gold nanoparticle. 

SiV color centers in diamond combined with diamond nanophotonic devices are proven to be an efficient platform for quantum optical networks \cite{sipahigil2016integrated,nguyen2019quantum}. Color centers in diamonds integrated with other photonic materials are investigated for quantum network applications \cite{schroder2016quantum,ruf2021quantum}. Single photons coupled to a single-mode fiber using SiV centers in a waveguide coupled to a diamond photonic crystal cavity have been demonstrated \cite{burek2017fiber}. A plasmonic waveguide integrated with a germanium-vacancy (GeV) center in nanodiamonds achieved $\eta$-value of $\sim$ 68\% by placing a Bragg reflector \cite{Siampour}. Also, gallium arsenide waveguides \cite{wang2014gallium} and indium phosphide nanotips \cite{reimer2012bright} have been demonstrated. 

Although these hybrid systems and modified SNWs enhance the $\eta$-value, it is challenging to introduce these structures into sub-wavelength waveguides. Hence a simple, easy-to-fabricate structure which can enhance the $\eta$-value draws attention. In this context, a pair consisting of these structures, which can enhance the $\eta$-value is at the cutting edge. Two parallel SNWs are found to enhance the $\eta_{p}$-value (coupling efficiency for a pair of nanostructures) \cite{le2020coupling,le2021spatial}. It has been numerically shown that by placing the single quantum emitter in the gap of two parallel SNWs, the $\eta_{p}$-value reaches $\sim$54\% \cite{shao2022twin}. A blade-like optical field from a higher dimensional cross section is generated using a pair of SNWs \cite{yang2024generating}. However, coupling efficiency for a pair of nanostructures consisting of SNTs has not been explored yet. Although research on coupling to DNWs is being pursued, detailed study on the coupling of photons into guided modes of the DNT has not been reported yet. 

In this paper, we numerically report highly efficient coupling of single photons from the SDS using a pair of nanostructures. The maximum $\eta_{p}$-value of 56\%, into guided modes of the SNT, is found when the SNT of radius 0.43 $\mu m$ is placed in the vicinity of the DNT and the DNW. Additionally, we found that varying the radius of the DNT/DNW does not significantly affect the $\eta_{p}$-value. Furthermore, we investigate the $\eta$-value of single photons from the SDS into guided modes of the DNT. The maximum $\eta$-value of 87\% is found for the radially oriented SDS on the facet of the DNT of radius 0.4 $\mu m$. We found that the $\eta_{p}$-value is enhanced when the DNT is placed in the vicinity of another DNT and the DNW.

\section{Simulation procedure}\label{Sim}
\begin{figure*}[h]
	\centering
	\includegraphics[width=0.9\textwidth]{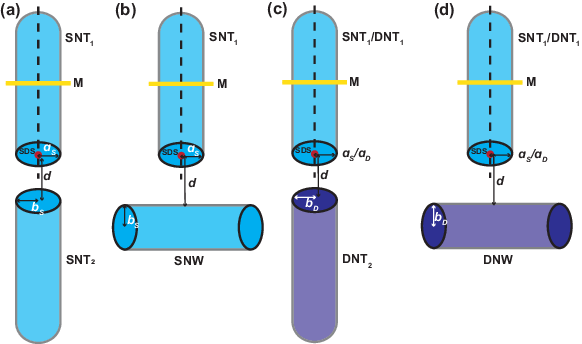}
	\caption{(a), (b), (c), and (d) show conceptual schematics for a silica nanotip (SNT$_{1}$) in the vicinity of another SNT, a silica nanowire, a diamond nanotip, and a diamond nanowire (DNW), respectively. (c) and (d) also shows conceptual schematics for a diamond nanotip (DNT$_{1}$) in the vicinity of another DNT and the DNW, respectively. SDS, M, \textit{a$_{S}$}, \textit{b$_{S}$}/\textit{b$_{D}$}, and \textit{d} denote single dipole source, monitor, radius of the SNT$_{1}$/DNT$_{1}$, radius of the second nanostructure, and distance of the second nanostructure from the SNT$_{1}$/DNT$_{1}$, respectively.}
	\label{fig1}
\end{figure*}

The finite difference time domain (FDTD) method (FDTD package, Ansys) is used to perform numerical simulations \cite{garcia2005new}. The conceptual schematics for coupling single photons into the SNT/DNT for different cases are shown in Figs. \ref{fig1} (a)-(d). The SNT$_{1}$/DNT$_{1}$ is a cylinder made of silica/diamond. The refractive index of the surrounding medium is set at 1, indicating vacuum. For quick simulations, the entire simulation area is set as $6$$\times$6$\times$$25$ $\mu m^3$. The SNT$_{1}$/DNT$_{1}$ extends along the $z$-axis, with a specified length of 25 $\mu m$. The SDS is positioned at the center of the facet of the SNT$_{1}$/DNT$_{1}$ and 10 $nm$ away, not in contact with the facet. The wavelength of the SDS is set at 620 $nm$, and orientation is set along the radial/axial direction of the SNT$_{1}$/DNT$_{1}$. Note that when the second nanostructure is SNW or DNW, the SDS orientation is set along the radial/azimuthal/axial direction. A 2D power monitor (M) is positioned at a distance of 15 $\mu m$ from the SDS to record the coupling power for the pair of nanostructures. The second SNT and DNT are parallel to the SNT$_{1}$ as shown in Figs. \ref{fig1} (a) and (c). The SNW and the DNW are positioned horizontally, as shown in Figs. \ref{fig1} (b) and (d).

\subsection{Coupling to guided modes of SNT}
\label{SNT}
The simulations are performed for the following four cases.\\

\noindent Case 1: SNT$_{1}$ in the vicinity of SNT$_{2}$\\
Case 2: SNT$_{1}$ in the vicinity of a SNW\\
Case 3: SNT$_{1}$ in the vicinity of DNT$_{2}$\\
Case 4: SNT$_{1}$ in the vicinity of a DNW\\

First, we perform simulations by varying the distance ($d$) between the SNT$_{1}$ and the second nanostructure from 0.05 $\mu m$ to 0.5 $\mu m$. All four cases are performed for three radii of the SNT$_{1}$- 0.20 $\mu m$ (P$_{1}$), 0.43 $\mu m$ (P$_{2}$), and 0.71 $\mu m$ (P$_{3}$). The radius of the second nanostructure is also set identical in each case. First we optimize the $d$-value for the maximum $\eta_{p}$-value, then the radius of the second nanostructure (\textit{b$_{D}$}) is varied.
\subsection{Coupling to guided modes of DNT}
First, we perform simulations to verify the coupling efficiency ($\eta$) of single photons into guided modes of the DNT. The radius of the DNT$_{1}$ is varied from 0.062 $\mu m$ to 1.24 $\mu m$. Optimizing the radius of the DNT$_{1}$ for the maximum $\eta$-value, simulations are performed to check the enhancement in the coupling to the DNT$_{1}$ in the vicinity of another DNT and the DNW. Other parameters and simulation procedures are as described in Sec. \ref{SNT}. The conceptual schematics are shown in Figs. \ref{fig1} (c) and (d). The simulations are performed for the following cases.\\

\noindent Case 5: SDS on the facet of the DNT$_{1}$\\
Case 6: DNT$_{1}$ in the vicinity of DNT$_{2}$\\
Case 7: DNT$_{1}$ in the vicinity of a DNW\\

\section{Results}
\begin{figure*}
	\centering
	\includegraphics[width=\textwidth]{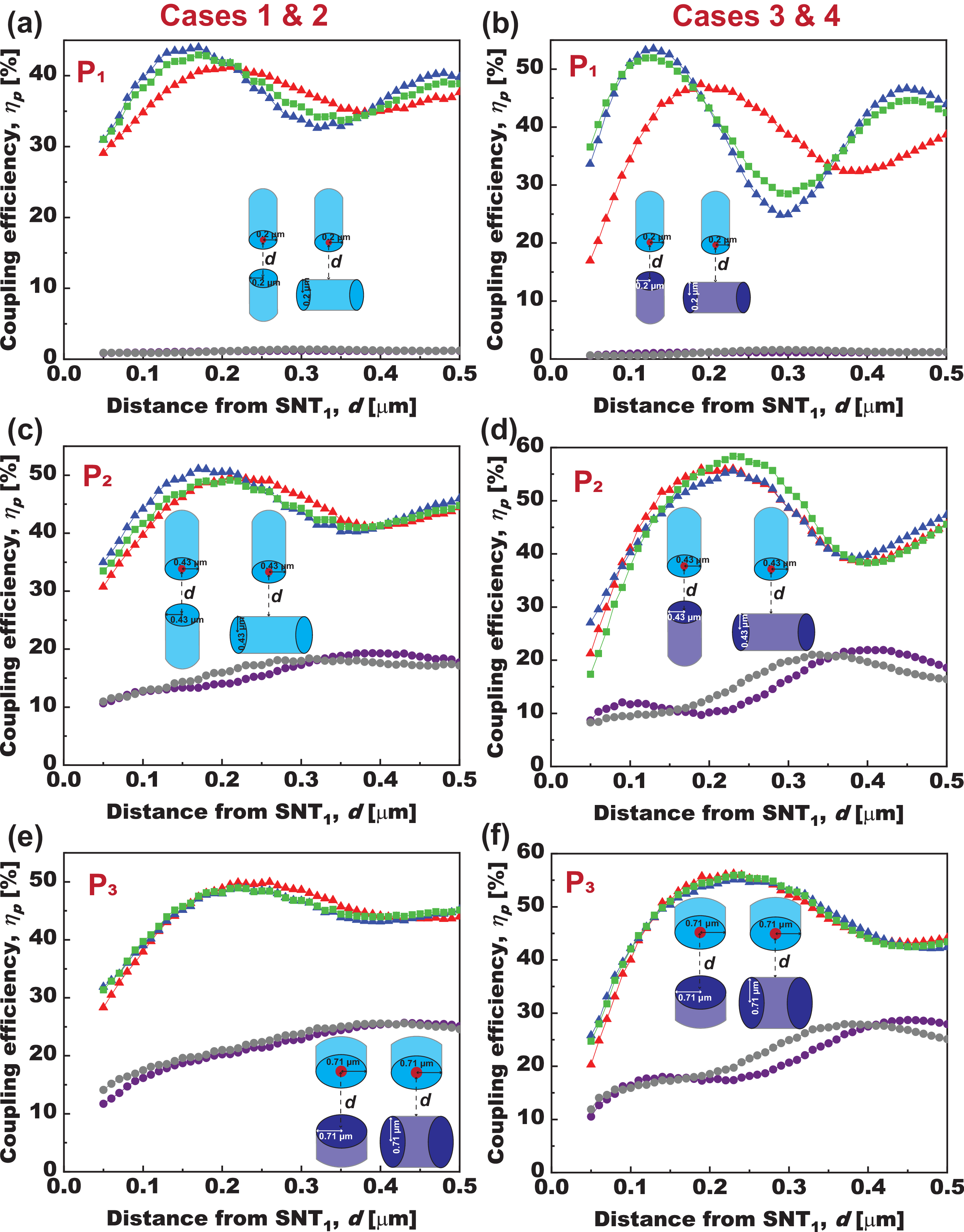}
	\caption{Dependence of coupling efficiency ($\eta_{p}$) as a function of distance from the silica nanotip (SNT$_{1}$) (\textit{d}) to the second nanostructure. For the SNT$_{1}$ of radius 0.2 $\mu m$: (a) SNT$_{2}$ and a silica nanowire (SNW), (b) a diamond nanotip (DNT$_{2}$) and a diamond nanowire (DNW). For SNT$_{1}$ of radius 0.43 $\mu m$: (c) SNT$_{2}$ and the SNW (d) the DNT$_{2}$ and the DNW. For SNT$_{1}$ of radius 0.71 $\mu m$: (e) SNT$_{2}$ and the SNW (f) the DNT$_{2}$ and the DNW. Red triangles and purple circles denote radially and axially oriented SDS, respectively, when the SNT$_{1}$ is in the vicinity of SNT$_{2}$/DNT$_{2}$. Blue triangles, green squares, and grey circles denote radially, azimuthally, and axially oriented SDS, respectively, when the SNT$_{1}$ is in the vicinity of the SNW/DNW. Insets show corresponding schematics.}
	\label{fig2}
\end{figure*}

The simulation predicted results for the four cases for the coupling into the SNT$_{1}$ corresponding to P$_{1}$, P$_{2}$, and P$_{3}$ are shown in Figs. \ref{fig2} (a)-(b), (c)-(d), and (e)-(f), respectively. Insets show corresponding schematics of the hybrid nanostructure. The radius of the nanostructures remains constant in each case, and the variable parameter is $d$-value only. The horizontal and vertical axes denote $d$-values and $\eta_{p}$-values, respectively. For Figs. \ref{fig2} (a)/(b), (c)/(d), and (e)/(f), red triangles and purple circles denote radially and axially oriented SDS, respectively, when the SNT$_{1}$ is in the vicinity of the SNT$_{2}$/DNT$_{2}$. Blue triangles, green squares, and grey circles denote radially, azimuthally, and axially oriented SDS, respectively when the SNT$_{1}$ is in the vicinity of the SNW/DNW. 

For P$_{1}$: In case 1, the maximum $\eta_{p}$-value ($\eta_{p}^{max}$) of 41\% occured at the $d$-value of 0.21 $\mu m$. In case 2, the $\eta_{p}^{max}$-value of 44\% occured at the $d$-value of 0.17 $\mu m$. In case 3, the $\eta_{p}^{max}$-value of 47\% occured at the $d$-value of 0.19 $\mu m$. In case 4, the $\eta_{p}^{max}$-value of 53.5\% occured at the $d$-value of 0.13 $\mu m$. All the $\eta_{p}^{max}$-value mentioned are for the radial orientation of the SDS. 

For P$_{2}$: In case 1, the $\eta_{p}^{max}$-value of 50\% occured at the $d$-value of 0.22 $\mu m$. In case 2, the $\eta_{p}^{max}$-value of 51\% occured at the $d$-value of 0.17 $\mu m$. In case 3, the $\eta_{p}^{max}$-value of 56\% occured at the $d$-value of 0.19 $\mu m$. In case 4, the $\eta_{p}^{max}$-value of 56\% occured at the $d$-value of 0.23 $\mu m$. All the $\eta_{p}^{max}$-value mentioned are for the radial orientation of the SDS.

For P$_{3}$: In case 1, the $\eta_{p}^{max}$-value of 50\% occured at the $d$-value of 0.22 $\mu m$. In case 2, the $\eta_{p}^{max}$-value of 49\% occured at the $d$-value of 0.22 $\mu m$. In case 3, the $\eta_{p}^{max}$-value of 56\% occured at the $d$-value of 0.23 $\mu m$. In case 4, the $\eta_{p}^{max}$-value of 55\% occured at the $d$-value of 0.23 $\mu m$. All the $\eta_{p}^{max}$-value mentioned are for the radial orientation of the SDS.

\begin{figure*}[h!]%
	\centering
	\includegraphics[width=\textwidth]{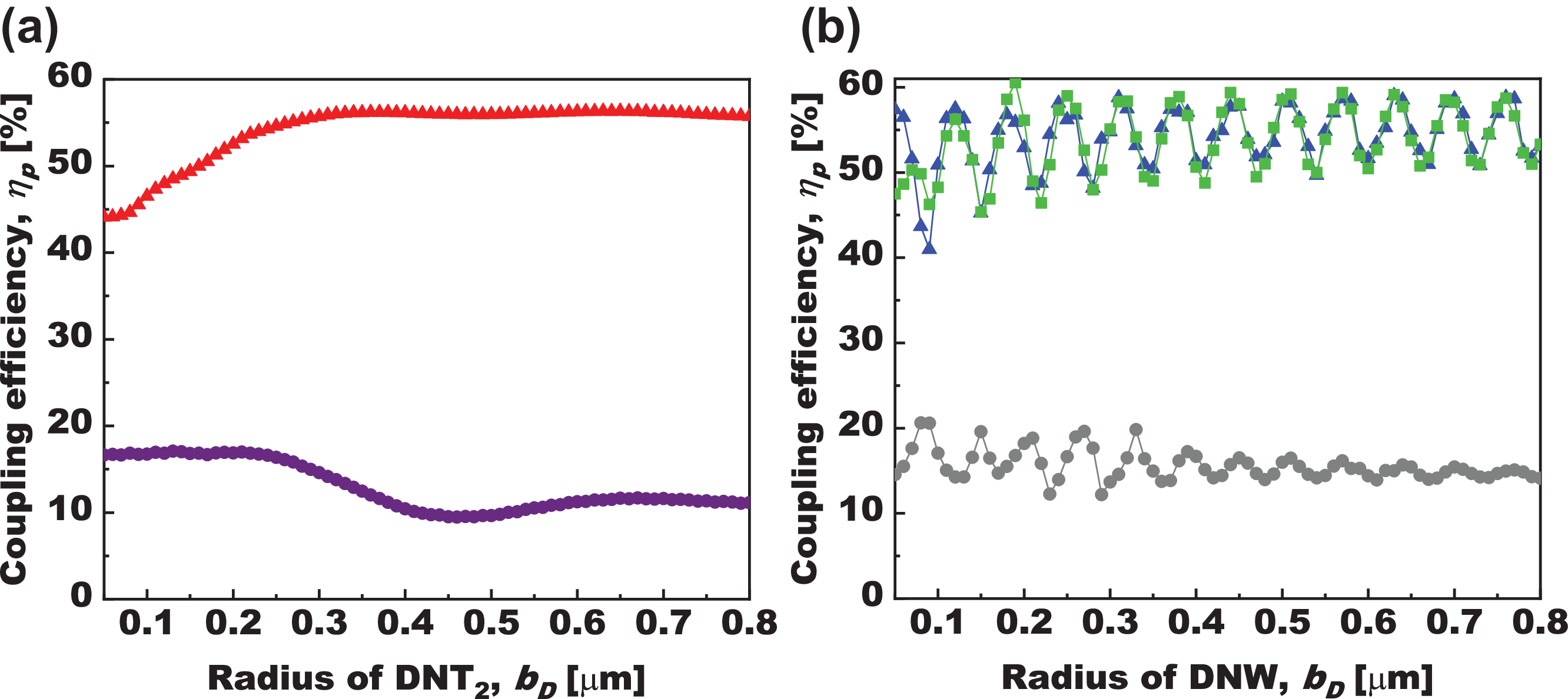}
	\caption{(a) and (b) show the dependence of coupling efficiency ($\eta_{p}$) as a function of the radius of the diamond nanotip and the diamond nanowire (\textit{b$_{D}$}), respectively, which are in the vicinity of the SNT$_{1}$. All the plots correspond to the SNT$_{1}$ radius of 0.43 $\mu m$. SNT denotes silica nanotip. Red/blue triangles, green squares, and purple/grey circles denote radially, azimuthally, and axially oriented SDS, respectively.}
	\label{fig3}
\end{figure*}

Next, we investigate the dependence of the $\eta_{p}$-value on the radius of the DNT$_{2}$ and the DNW \textit{(b$_{D}$)}. The radius of the SNT$_{1}$ is set at 0.43 $\mu m$. The distance between the SNT$_{1}$ and the DNT$_{2}$/DNW is set at 0.19 $\mu m$/0.23 $\mu m$. The dependence of $\eta_{p}$-value on the radius of the DNT$_{2}$ and the DNW is shown in Figs. \ref{fig3} (a) and (b), respectively. The horizontal and vertical axes denote \textit{b$_{D}$}-values and $\eta_{p}$-values, respectively. Red/blue triangles, green squares, and purple/grey circles denote radially, azimuthally, and axially oriented SDS, respectively. The $\eta_{p}$-value is almost constant with change in the radius of the DNT$_{2}$ from 0.3 $\mu m$ to 0.8 $\mu m$. The $\eta_{p}$-value slightly decreases and then increases with change in the radius of the DNW.

\begin{figure*}[h!]%
	\centering
	\includegraphics[width=\linewidth]{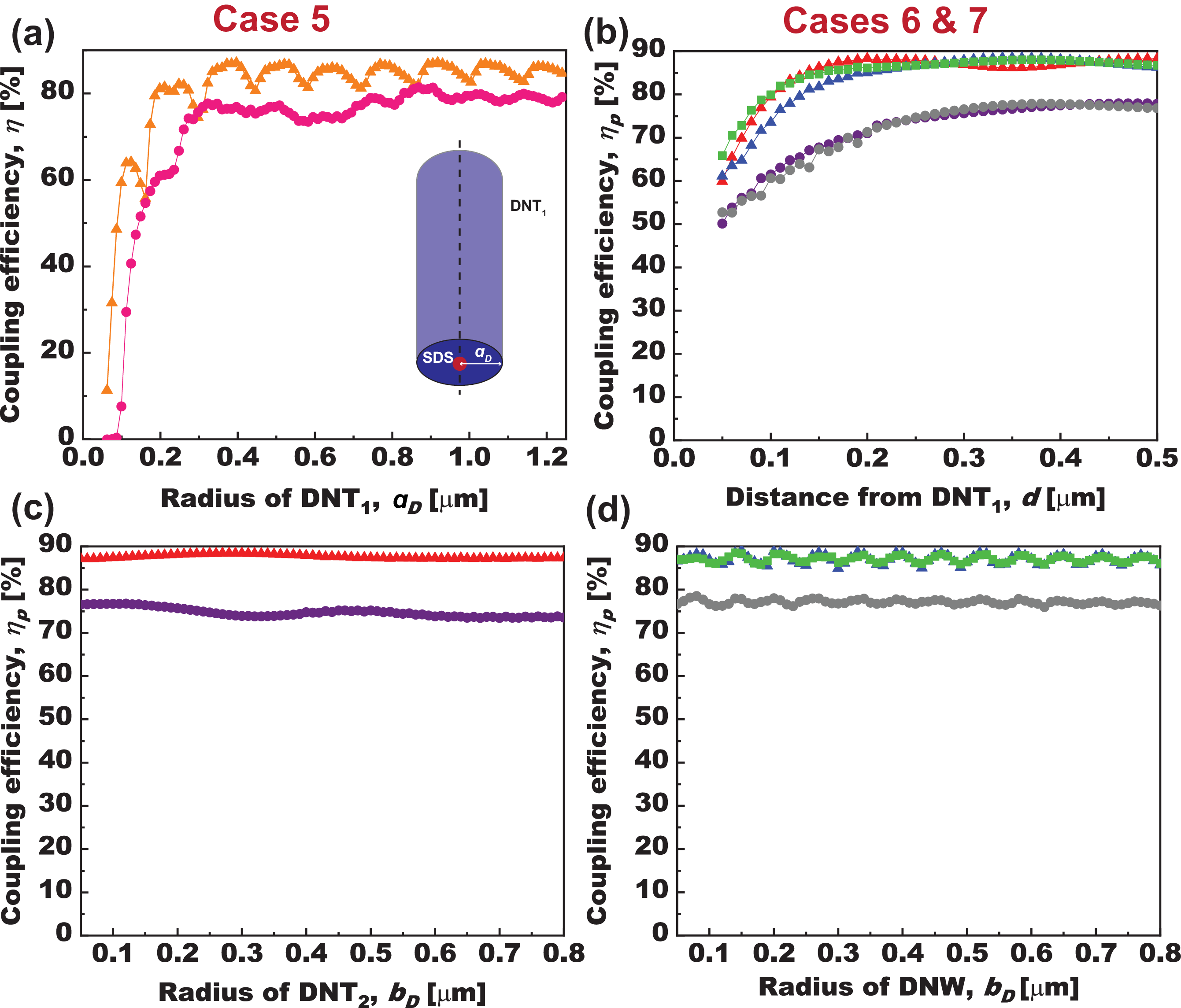}
	\caption{(a) Dependence of coupling efficiency ($\eta$) on the radius of the diamond nanotip (DNT$_{1}$) (\textit{a}) for the single dipole source (SDS) placed on the facet of it. Orange triangles and pink circles denote radially and axially oriented SDS, respectively. Inset shows a conceptual schematic for the SDS on the facet of the DNT$_{1}$. (b) Dependence of coupling efficiency ($\eta_{p}$) as a function of distance from the DNT$_{1}$ (\textit{d}) to DNT$_{2}$ and a diamond nanowire (DNW). Red triangles and purple circles denote radially and axially oriented SDS, respectively when the DNT$_{1}$ is in the vicinity of DNT$_{2}$. Blue triangles, green squares, and grey circles denote radially, azimuthally, and axially oriented SDS, respectively when the DNT$_{1}$ is in the vicinity of the DNW. (c) and (d) show dependence of $\eta_{p}$-value as a function of radius of the DNT$_{2}$ and the DNW (\textit{b$_{D}$}), respectively. (c) and (d) correspond to the DNT$_{1}$ radius of 0.4 $\mu m$ }
	\label{fig4}
\end{figure*}

The simulation predicted results for the coupling into guided modes of the DNT$_{1}$ (case 5) is shown in Fig. \ref{fig4} (a). The horizontal axis denotes the radius of the DNT$_{1}$. The vertical axis denotes $\eta$-value. Orange triangles and pink circles denote radially and axially oriented SDS, respectively. The maximum $\eta$-value of 87\% occured at the DNT$_{1}$ radius of 0.4 $\mu m$. 

The simulation predicted results for cases 6 \& 7,corresponding to the DNT$_{1}$ radius of 0.4 $\mu m$, is shown in Fig. \ref{fig4} (b). The horizontal and vertical axes denote $d$-values and $\eta_{p}$-values, respectively. Red triangles and purple circles denote radially and axially oriented SDS, respectively when the DNT$_{1}$ is in the vicinity of DNT$_{2}$. Blue triangles, green squares, and grey circles denote radially, azimuthally, and axially oriented SDS, respectively when the DNT$_{1}$ is in the vicinity of the DNW. For the case of two DNTs, the $\eta_{p}^{max}$-value of 88\% occured at the $d$-value of 0.2 $\mu m$. For the case of the DNT and the DNW, the $\eta_{p}^{max}$-value of 88.5\% occured at the $d$-value of 0.35 $\mu m$. 

Next, the distance between the DNT$_{1}$ and the DNT$_{2}$/DNW is set at 0.2 $\mu m$/0.35 $\mu m$. The radius of the DNT$_{2}$/DNW (\textit{b$_{D}$}) is varied from 0.05 $\mu m$ to 0.8 $\mu m$. The dependence of $\eta_{p}$-value on the radius of the DNT$_{2}$ and the DNW are shown in Figs. \ref{fig4} (c) and (d), respectively. The horizontal and vertical axes denote $b_{D}$-values and $\eta_{p}$-values, respectively. The $\eta_{p}$-value is almost constant with change in the radius of the DNT$_{2}$. The $\eta_{p}$-value slightly decreases and then increases with change in the radius of the DNW.
\newpage
\section{Discussions}\label{Disc}

\begin{table}[htb]
	\centering
	\caption{Summary of maximum coupling efficienices}\label{tab1}%
	\begin{tabular}{|c|c|c|c|c|c|c|c|c|c|c|c|}
		\hline
		
		\multicolumn{1}{|c|}{S.No.}& \multicolumn{1}{|c|}{Case}&\multicolumn{9}{c|}{Maximum coupling efficiency} \\ \hline
		
		\multicolumn{1}{|c|}{} & \multicolumn{1}{|c|}{} & \multicolumn{3}{|c|}{P$_1$} & \multicolumn{3}{c|}{P$_2$} & \multicolumn{3}{c|}{P$_3$} \\ \hline
		\multicolumn{1}{|c|}{}& \multicolumn{1}{|c|}{}& $\eta^{max}$ & $d$& $r$&$\eta^{max}$ & $d$ & $r$& $\eta^{max}$ & $d$& $r$\\
		\multicolumn{1}{|c|}{}& \multicolumn{1}{|c|}{}& &($\mu m$)&($\mu m$)&&($\mu m$) &($\mu m$) &&($\mu m$) &($\mu m$)\\
		\hline
		1 & SNT,& 37\% & & 0.2 & 43.5\% & & 0.43 & 44\% & & 0.71\\ 
		 &SDS \cite{resmi2023efficient,resmichanneling} & & && & & && &\\ 
		2 & Case 1 & 41\%  & 0.21 & 0.2 & 50\% & 0.22 & 0.43 & 50\% & 0.22 & 0.71 \\
		3 & Case 2 & 44\%  & 0.17& 0.2  & 51\% & 0.17& 0.43& 49\% & 0.22 & 0.71  \\
		4 & Case 3 & 47\%  & 0.19 & 0.2 & 56\% & 0.19& 0.43 & 56\% & 0.23 & 0.71  \\
		5 & Case 4 & 53.5\%  & 0.13& 0.2& 56\% & 0.23 & 0.43 & 55\% & 0.23 & 0.71\\
		6 & Case 5 & 64\%  & & 0.12& 82\% & & 0.25 & 87\% & & 0.4\\ \hline
		\multicolumn{1}{|c|}{7} &\multicolumn{1}{|c|}{Case 6} & \multicolumn{9}{c|}{88\%}\\
		\multicolumn{1}{|c|}{8} &\multicolumn{1}{|c|}{Case 7} & \multicolumn{9}{c|}{88.5\%}\\ \hdashline
		\multicolumn{1}{|c|}{9} &\multicolumn{1}{|c|}{SNW, Single} & \multicolumn{9}{c|}{28\%}\\
		\multicolumn{1}{|c|}{} &\multicolumn{1}{|c|}{atom \cite{le2005spontaneous}} & \multicolumn{9}{c|}{}\\
		\multicolumn{1}{|c|}{10} &\multicolumn{1}{|c|}{SNW, QD \cite{yalla2012efficient}} & \multicolumn{9}{c|}{22\%}\\
		\multicolumn{1}{|c|}{11} &\multicolumn{1}{|c|}{SNT, SDS  \cite{chonan2014efficient}} & \multicolumn{9}{c|}{38\%}\\
		\multicolumn{1}{|c|}{12} &\multicolumn{1}{|c|}{DNT, SDS \cite{das2023efficient}} & \multicolumn{9}{c|}{61\%}\\
		\multicolumn{1}{|c|}{13} & \multicolumn{1}{|c|}{SNW, Bragg} & \multicolumn{9}{c|}{$\sim$84\%}\\
		\multicolumn{1}{|c|}{} & \multicolumn{1}{|c|}{cavity \cite{schell2015highly,takashima2016detailed}} & \multicolumn{9}{c|}{}\\
		\multicolumn{1}{|c|}{14} &\multicolumn{1}{|c|}{SNW,} & \multicolumn{9}{c|}{16-37\%}\\
		\multicolumn{1}{|c|}{} &\multicolumn{1}{|c|}{NV centre \cite{patel2016efficient}} & \multicolumn{9}{c|}{}\\
		\multicolumn{1}{|c|}{15} & \multicolumn{1}{|c|}{SNW,} & \multicolumn{9}{c|}{$\sim$75\%}\\
		\multicolumn{1}{|c|}{} & \multicolumn{1}{|c|}{NV centre \cite{yonezu2017efficient}} & \multicolumn{9}{c|}{}\\
		\multicolumn{1}{|c|}{16} & \multicolumn{1}{|c|}{SNW, Groove} & \multicolumn{9}{c|}{80\%}\\
		\multicolumn{1}{|c|}{} & \multicolumn{1}{|c|}{in cavity \cite{li2018tailoring}} & \multicolumn{9}{c|}{}\\
		\multicolumn{1}{|c|}{17} & \multicolumn{1}{|c|}{SNW,} & \multicolumn{9}{c|}{85\%}\\
		\multicolumn{1}{|c|}{} & \multicolumn{1}{|c|}{cavity \cite{nayak2019real}} & \multicolumn{9}{c|}{}\\
		\multicolumn{1}{|c|}{18} & \multicolumn{1}{|c|}{SNW,} & \multicolumn{9}{c|}{$\sim$63\%}\\
		\multicolumn{1}{|c|}{} & \multicolumn{1}{|c|}{Hole \cite{wang2021high}} & \multicolumn{9}{c|}{}\\
		\multicolumn{1}{|c|}{19} & \multicolumn{1}{|c|}{SNW,} & \multicolumn{9}{c|}{$\sim$4\%}\\
		\multicolumn{1}{|c|}{} & \multicolumn{1}{|c|}{SiV \cite{yalla2022integration}} & \multicolumn{9}{c|}{}\\
		\multicolumn{1}{|c|}{20} & \multicolumn{1}{|c|}{SNT,} & \multicolumn{9}{c|}{$\sim$43\%}\\
		\multicolumn{1}{|c|}{} & \multicolumn{1}{|c|}{gold \cite{das2023efficient}} & \multicolumn{9}{c|}{}\\
		\multicolumn{1}{|c|}{21} & \multicolumn{1}{|c|}{DNT,} & \multicolumn{9}{c|}{$\sim$64\%}\\
		\multicolumn{1}{|c|}{} & \multicolumn{1}{|c|}{gold \cite{das2023efficient}} & \multicolumn{9}{c|}{}\\
		\multicolumn{1}{|c|}{22} & \multicolumn{1}{|c|}{Parallel} & \multicolumn{9}{c|}{$\sim$54\%}\\
		\multicolumn{1}{|c|}{} & \multicolumn{1}{|c|}{SNWs \cite{shao2022twin}} & \multicolumn{9}{c|}{}\\
		\multicolumn{1}{|c|}{23} & \multicolumn{1}{|c|}{Dielectric, Bragg} & \multicolumn{9}{c|}{$\sim$68\%}\\
		\multicolumn{1}{|c|}{} & \multicolumn{1}{|c|}{reflector \cite{Siampour}} & \multicolumn{9}{c|}{}\\
		\hline
	\end{tabular}
\end{table}

The summary of maximum coupling efficiencies of the present cases and other results available in the literature is shown in Table \ref{tab1}. The first row shows the result for the SNT and the SDS system. The second to eighth rows are the present results, and the rest are results for other reported systems. The ninth to twelfth rows are for quantum emitters on a single structure, and the thirteenth to twenty-third rows are for quantum emitters on a hybrid structure consisting of cavities or plasmonic particles.

As we reported earlier \cite{resmi2023efficient}, the $\eta$- value without a second nanostructure for P$_{1}$ is 37\%. As seen in Table \ref{tab1}, the $\eta_{p}^{max}$-value for  P$_{1}$ is 41\% in the vicinity of the SNT$_{2}$. One can readily see that when the SNT$_{2}$ is replaced with the SNW, $\eta_{p}$-value increases from 41\% to 44\%. Further it enhances to 47\% when the SNT$_{2}$ is replaced with the DNT$_{2}$. It is inferred that the $\eta_{p}$-value may enhance more when the DNW is used. As seen in Table \ref{tab1}, the $\eta_{p}$-value increased to 53.5\%. This may be due to the reflection area of the SNW/DNW is higher than that of the SNT$_{2}$/DNT$_{2}$. As the reflection increases, $\eta_{p}$-value also increases.

Compared to P$_{1}$, the $\eta_{p}$-value for  P$_{2}$ is more as seen in Table \ref{tab1}. SNT$_{1}$ in the vicinity of SNT$_{2}$ achieves $\eta_{p}^{max}$-value of 50\% and increses to 56\% in the vicinity of the DNT$_{2}$. It is clear from the results that when the second nanostructure of a higher refractive index material is placed, the $\eta_{p}$-value is enhanced. One can readily see in Fig. \ref{fig2} (d) that the $\eta_{p}$-value corresponding to the azimuthal orientation is slightly greater than that of the radial orientation when the SNT$_{1}$ is in the vicinity of the DNW. The area where the SDS interacts with the DNW is more in azimuthal orientation (parallel to the DNW), corresponding to more reflection area and hence a higher $\eta_{p}$-value. The $\eta_{p}^{max}$-value corresponding to the perpendicular/parallel orientation of the SDS with respect to the DNW is 56\% / 58\%. 

As seen in Table \ref{tab1}, the $\eta_{p}^{max}$-value for the four different cases corresponding to P$_{3}$ are similar to that of  P$_{2}$. Hence, considering a smaller radius, P$_{2}$ is optimized to determine the $\eta_{p}$-value dependence on the radius of the second nanostructure. One can readily see from the results that the $\eta_{p}$-value is enhanced when the SNT$_{1}$ is in the vicinity of the DNT$_{2}$ and the DNW. Also, a slight variation in the \textit{d}-value does not affect the $\eta_{p}$-value.

As seen in Fig. \ref{fig3} (a), the $\eta_{p}$-value does not vary significantly when the radius of the DNT$_{2}$ is varied from 0.3 $\mu m$ to 0.8 $\mu m$. It indicates that the DNT$_{2}$ of a specific radius is not required for corresponding experiments. Hence, the experimental freedom of fabricating the DNT of radius in a wide range exists. The difficulty in fabricating the SNT and the DNT identical in radius can be avoided. However, for the radius from 0.3 $\mu m$ to 0.05 $\mu m$, the $\eta_{p}$-value can be found decreasing. The solid angle of the SNT$_{1}$ may not be covered by the DNT$_{2}$ of radius less than 0.3 $\mu m$. As seen in Fig. \ref{fig3} (b), although the $\eta_{p}$-value slightly varies with respect to the DNW radius, it oscillates between 50\% and 58\%. It indicates that the radius of the DNW may not necessarily be a specific value.

The coupling of single photons from the SDS into guided modes of the DNT$_{1}$ as shown in Table \ref{tab1}, indicates that a maximum $\eta$-value of 87\% is achieved for the radius of 0.4 $\mu m$. This is due to the higher refractive index of diamond. The collection material of higher refractive index implies stronger light confinement and thus coupling is higher. Although using a pair of structures for the SNT, the $\eta_{p}^{max}$-value could not cross 60\%, but the single DNT achieves the $\eta$-value of 87\%. This result is highly significant for applications in quantum technology and computation. As seen in Table. \ref{tab1}, the DNT$_{1}$ in the vicinity of DNT$_{2}$ and the DNW attains the $\eta_{p}^{max}$-value of 88\%. A significant enhancement in the $\eta$-value is not observed here in contrast to the case of the SNT since $\eta$-value is saturated already. Although if a pair of structures have to be used, the radius of the second nanostructure does not affect the $\eta_{p}$-value as shown in Figs. \ref{fig4} (c) and (d). For the case of the DNT$_{2}$, the $\eta_{p}$-value does not vary significantly. For the case of the DNW, the $\eta_{p}$-value oscillates between 84\% and 88\%. Hence, the experimental difficulty in fabricating two identical DNTs is minimized.

As shown in Table \ref{tab1}, one can readily observe the significant $\eta$-value of the present system compared to other reported results. For a single atom on the SNW \cite{le2005spontaneous} and for a single QD on the SNW \cite{yalla2012efficient}, the $\eta^{max}$ is 28\% and 22\%, respectively. But using the SNT can give higher $\eta$-value as per our simulation results. Though the SDS on the SNT in the single mode regime \cite{chonan2014efficient} attains the $\eta^{max}$ of 38\%, it is more in the multimode regime \cite{resmi2023efficient,resmichanneling}. The $\eta^{max}$-value of 61\% for the SDS on the facet of the DNT in the single mode regime is reported \cite{das2023efficient}. But our results show that $\eta^{max}$-value of 87\% is possible in multimode regime. Bragg cavity on the SNW achieves the $\eta$-value upto 84\% \cite{schell2015highly,takashima2016detailed}. The hybrid system consisting of the SNW and a nitrogen vacancy (NV) centre on the DNW predicts the maximum $\eta$-value of 75\% \cite{patel2016efficient}. A narrow air-filled groove into the cavity on the SNW attains the maximum $\eta$-value of 80\% \cite{li2018tailoring}. The maximum $\eta$-value of 85\% is obtained for single atoms interfaced on photonic crystal structures fabricated on the SNW \cite{nayak2019real}. Single quantum emitter placed in a hole etched on the SNW can enhnace the $\eta$-value up to $\sim$63\% \cite{wang2021high}. A Bragg reflector placed near the dielectric-GeV center system achieves the $\eta$-value of $\sim$68\% \cite{Siampour}. Though these results are comparable with the present result, a single diamond nanostructure achieving the maximum $\eta$-value of 87\% is the prominent factor. Thus the results presented here are on par with systems that needed gratings or other hybrid structures to be introduced on the SNW and the SNT.

Regarding the SNT, an increase in the $\eta$-value up to 43\% is obtained when a gold nanoparticle is in the proximity to it \cite{das2023efficient}. However, the SNT in the vicinity of the second nanostructure, as we described, achieves a higher $\eta$-value. The $\eta$-value for the DNT and the SDS is enhanced to 64\% in the presence of a gold nanoparticle \cite{das2023efficient} in the single mode regime. However, as per our results, it can be inferred that the DNT in the vicinity of another DNT/DNW might produce a higher $\eta$-value. The $\eta^{max}$-value for the SDS in the gap of two parallel SNWs is 54\% \cite{shao2022twin}. But the SNT and the DNT/DNW system offers higher $\eta$-value or a single DNT can be used for enhanced $\eta$-value. The $\eta^{max}$-value of 87\% can be further enhanced to unity if we incorporate cavity on the DNT. Also, a Bragg reflector can be placed on one side of the present system to enhance the $\eta$-value. \cite{Siampour}.

The advantage of the present result is the ease of fabricating the structure. The added advantage is the freedom of having the second nanostructure's radius in a wide range. The SNT can be fabricated in an easy and cost-effectively by chemical etching technique \cite{resmifabrication}. The experimental feasibility of coupling a single photon source on the SNT is confirmed using QDs \cite{resmichanneling}. Hence, a pair of nanostructures consisting of the SNT is feasible. The fabrication of the DNT and the DNW is experimentally feasible as diamond optomechanical crystals are fabricated using angled etching technique \cite{burek2012free} and are proven for quantum optical applications \cite{burek2016diamond}. The placement of the second nanostructure is possible using a nanopositioning stage.

\section{Conclusion}\label{Con}
In conclusion, we numerically reported highly efficient coupling of single photons from a single dipole source (SDS) using a pair of nanostructures. The maximum coupling efficiency ($\eta_{p}$) of 56\%, into guided modes of a silica nanotip (SNT), was found when the SNT of radius 0.43 $\mu m$ was placed in the vicinity of a diamond nanotip (DNT) and a diamond nanowire (DNW). Additionally, we found that varying the radius of the DNT/DNW does not significantly affect the $\eta_{p}$-value. This ensures experimental freedom of fabricating the DNT of radius in a wide range. Furthermore, we investigated the coupling of single photons from the SDS into guided modes of the DNT. The maximum coupling efficiency ($\eta$) of 87\% was found when the radially oriented SDS was placed on the facet of the DNT of radius 0.4 $\mu m$. We found that the $\eta_{p}$-value was enhanced when the DNT was placed in the vicinity of another DNT and the DNW. The present results may open new possibilities in quantum networks. The $\eta$-value can be further enhanced by incorporating cavity or a Bragg reflector.

\backmatter
\bmhead{Acknowledgments}
RM acknowledges the University Grants Commission (UGC) for the financial support (Ref. No.:1412/CSIR-UGC NET June 2019). SS acknowledges funding support for the Chanakya-PG fellowship from the National Mission on Interdisciplinary Cyber-Physical Systems, of the Department of Science and Technology, Govt. of India through the I-HUB Quantum Technology Foundation (File No. I-HUB/PGF/2022-23/01). RRY acknowledges support from the Institute of Eminence (IoE) grant at the University of Hyderabad (File No. RC2-21-019) and the Science and Engineering Research Board (SERB) for the Core Research Grant (CRG) (File No. CRG/2021/009185).
\bmhead{Author Contributions}

RM performed simulations and analysed data.
\bmhead{Data availability}
Data presented in this paper are available from the authors upon reasonable request.

\section*{Declarations}
\textbf{Conflict of interest.} The authors declare no conflict of interests.\\
\textbf{Competing interests.} The authors declare no competing interests.\\
\textbf{Data availability.} Data may be obtained from authors upon reasonable request.\\
\textbf{Ethical approval.} Not applicable. No human and/or animal studies have been executed.

\bibliographystyle{sn-mathphys.bst}
\bibliography{sn-bibliography}


\begin{thebibliography}{45}
\ifx \bisbn   \undefined \def \bisbn  #1{ISBN #1}\fi
\ifx \binits  \undefined \def \binits#1{#1}\fi
\ifx \bauthor  \undefined \def \bauthor#1{#1}\fi
\ifx \batitle  \undefined \def \batitle#1{#1}\fi
\ifx \bjtitle  \undefined \def \bjtitle#1{#1}\fi
\ifx \bvolume  \undefined \def \bvolume#1{\textbf{#1}}\fi
\ifx \byear  \undefined \def \byear#1{#1}\fi
\ifx \bissue  \undefined \def \bissue#1{#1}\fi
\ifx \bfpage  \undefined \def \bfpage#1{#1}\fi
\ifx \blpage  \undefined \def \blpage #1{#1}\fi
\ifx \burl  \undefined \def \burl#1{\textsf{#1}}\fi
\ifx \doiurl  \undefined \def \doiurl#1{\url{https://doi.org/#1}}\fi
\ifx \betal  \undefined \def \betal{\textit{et al.}}\fi
\ifx \binstitute  \undefined \def \binstitute#1{#1}\fi
\ifx \binstitutionaled  \undefined \def \binstitutionaled#1{#1}\fi
\ifx \bctitle  \undefined \def \bctitle#1{#1}\fi
\ifx \beditor  \undefined \def \beditor#1{#1}\fi
\ifx \bpublisher  \undefined \def \bpublisher#1{#1}\fi
\ifx \bbtitle  \undefined \def \bbtitle#1{#1}\fi
\ifx \bedition  \undefined \def \bedition#1{#1}\fi
\ifx \bseriesno  \undefined \def \bseriesno#1{#1}\fi
\ifx \blocation  \undefined \def \blocation#1{#1}\fi
\ifx \bsertitle  \undefined \def \bsertitle#1{#1}\fi
\ifx \bsnm \undefined \def \bsnm#1{#1}\fi
\ifx \bsuffix \undefined \def \bsuffix#1{#1}\fi
\ifx \bparticle \undefined \def \bparticle#1{#1}\fi
\ifx \barticle \undefined \def \barticle#1{#1}\fi
\bibcommenthead
\ifx \bconfdate \undefined \def \bconfdate #1{#1}\fi
\ifx \botherref \undefined \def \botherref #1{#1}\fi
\ifx \url \undefined \def \url#1{\textsf{#1}}\fi
\ifx \bchapter \undefined \def \bchapter#1{#1}\fi
\ifx \bbook \undefined \def \bbook#1{#1}\fi
\ifx \bcomment \undefined \def \bcomment#1{#1}\fi
\ifx \oauthor \undefined \def \oauthor#1{#1}\fi
\ifx \citeauthoryear \undefined \def \citeauthoryear#1{#1}\fi
\ifx \endbibitem  \undefined \def \endbibitem {}\fi
\ifx \bconflocation  \undefined \def \bconflocation#1{#1}\fi
\ifx \arxivurl  \undefined \def \arxivurl#1{\textsf{#1}}\fi
\csname PreBibitemsHook\endcsname

\bibitem[\protect\citeauthoryear{Kimble}{2008}]{kimble2008quantum}
\begin{barticle}
\bauthor{\bsnm{Kimble}, \binits{H.J.}}:
\batitle{The quantum internet}.
\bjtitle{Nature}
\bvolume{453}(\bissue{7198}),
\bfpage{1023}--\blpage{1030}
(\byear{2008})
\end{barticle}
\endbibitem

\bibitem[\protect\citeauthoryear{O'brien et~al.}{2009}]{o2009photonic}
\begin{barticle}
\bauthor{\bsnm{O'brien}, \binits{J.L.}},
\bauthor{\bsnm{Furusawa}, \binits{A.}},
\bauthor{\bsnm{Vu{\v{c}}kovi{\'c}}, \binits{J.}}:
\batitle{Photonic quantum technologies}.
\bjtitle{Nature Photonics}
\bvolume{3}(\bissue{12}),
\bfpage{687}--\blpage{695}
(\byear{2009})
\end{barticle}
\endbibitem

\bibitem[\protect\citeauthoryear{O'brien}{2007}]{o2007optical}
\begin{barticle}
\bauthor{\bsnm{O'brien}, \binits{J.L.}}:
\batitle{Optical quantum computing}.
\bjtitle{Science}
\bvolume{318}(\bissue{5856}),
\bfpage{1567}--\blpage{1570}
(\byear{2007})
\end{barticle}
\endbibitem

\bibitem[\protect\citeauthoryear{Aharonovich
  et~al.}{2016}]{aharonovich2016solid}
\begin{barticle}
\bauthor{\bsnm{Aharonovich}, \binits{I.}},
\bauthor{\bsnm{Englund}, \binits{D.}},
\bauthor{\bsnm{Toth}, \binits{M.}}:
\batitle{Solid-state single-photon emitters}.
\bjtitle{Nature photonics}
\bvolume{10}(\bissue{10}),
\bfpage{631}--\blpage{641}
(\byear{2016})
\end{barticle}
\endbibitem

\bibitem[\protect\citeauthoryear{Kiraz et~al.}{2004}]{kiraz2004quantum}
\begin{barticle}
\bauthor{\bsnm{Kiraz}, \binits{A.}},
\bauthor{\bsnm{Atat{\"u}re}, \binits{M.}},
\bauthor{\bsnm{Imamo{\u{g}}lu}, \binits{A.}}:
\batitle{Quantum-dot single-photon sources: Prospects for applications in
  linear optics quantum-information processing}.
\bjtitle{Physical Review A}
\bvolume{69}(\bissue{3}),
\bfpage{032305}
(\byear{2004})
\end{barticle}
\endbibitem

\bibitem[\protect\citeauthoryear{Badolato
  et~al.}{2005}]{badolato2005deterministic}
\begin{barticle}
\bauthor{\bsnm{Badolato}, \binits{A.}},
\bauthor{\bsnm{Hennessy}, \binits{K.}},
\bauthor{\bsnm{Atature}, \binits{M.}},
\bauthor{\bsnm{Dreiser}, \binits{J.}},
\bauthor{\bsnm{Hu}, \binits{E.}},
\bauthor{\bsnm{Petroff}, \binits{P.M.}},
\bauthor{\bsnm{Imamoglu}, \binits{A.}}:
\batitle{Deterministic coupling of single quantum dots to single nanocavity
  modes}.
\bjtitle{Science}
\bvolume{308}(\bissue{5725}),
\bfpage{1158}--\blpage{1161}
(\byear{2005})
\end{barticle}
\endbibitem

\bibitem[\protect\citeauthoryear{Solomon et~al.}{2001}]{solomon2001single}
\begin{barticle}
\bauthor{\bsnm{Solomon}, \binits{G.}},
\bauthor{\bsnm{Pelton}, \binits{M.}},
\bauthor{\bsnm{Yamamoto}, \binits{Y.}}:
\batitle{Single-mode spontaneous emission from a single quantum dot in a
  three-dimensional microcavity}.
\bjtitle{Physical Review Letters}
\bvolume{86}(\bissue{17}),
\bfpage{3903}
(\byear{2001})
\end{barticle}
\endbibitem

\bibitem[\protect\citeauthoryear{Hood et~al.}{2000}]{hood2000atom}
\begin{barticle}
\bauthor{\bsnm{Hood}, \binits{C.J.}},
\bauthor{\bsnm{Lynn}, \binits{T.}},
\bauthor{\bsnm{Doherty}, \binits{A.}},
\bauthor{\bsnm{Parkins}, \binits{A.}},
\bauthor{\bsnm{Kimble}, \binits{H.}}:
\batitle{The atom-cavity microscope: Single atoms bound in orbit by single
  photons}.
\bjtitle{Science}
\bvolume{287}(\bissue{5457}),
\bfpage{1447}--\blpage{1453}
(\byear{2000})
\end{barticle}
\endbibitem

\bibitem[\protect\citeauthoryear{Elaganuru et~al.}{2024}]{bashachanneling}
\begin{barticle}
\bauthor{\bsnm{Elaganuru}, \binits{B.}},
\bauthor{\bsnm{Resmi}, \binits{M.}},
\bauthor{\bsnm{Ramachandrarao}, \binits{Y.}}:
\batitle{Highly efficient channeling of single photons into guided modes of
  optical nanocapillary fibers}.
\bjtitle{Optical and Quantum Electronics}
\bvolume{56}(\bissue{5}),
\bfpage{893}
(\byear{2024})
\end{barticle}
\endbibitem

\bibitem[\protect\citeauthoryear{Hausmann et~al.}{2013}]{hausmann2013coupling}
\begin{barticle}
\bauthor{\bsnm{Hausmann}, \binits{B.J.M.}},
\bauthor{\bsnm{Shields}, \binits{B.J.}},
\bauthor{\bsnm{Quan}, \binits{Q.}},
\bauthor{\bsnm{Chu}, \binits{Y.}},
\bauthor{\bsnm{Leon}, \binits{N.P.}},
\bauthor{\bsnm{Evans}, \binits{R.}},
\bauthor{\bsnm{Burek}, \binits{M.J.}},
\bauthor{\bsnm{Zibrov}, \binits{A.S.}},
\bauthor{\bsnm{Markham}, \binits{M.}},
\bauthor{\bsnm{Twitchen}, \binits{D.}}, \betal:
\batitle{Coupling of nv centers to photonic crystal nanobeams in diamond}.
\bjtitle{Nano letters}
\bvolume{13}(\bissue{12}),
\bfpage{5791}--\blpage{5796}
(\byear{2013})
\end{barticle}
\endbibitem

\bibitem[\protect\citeauthoryear{Wu and Tong}{2013}]{wu2013optical}
\begin{barticle}
\bauthor{\bsnm{Wu}, \binits{X.}},
\bauthor{\bsnm{Tong}, \binits{L.}}:
\batitle{Optical microfibers and nanofibers}.
\bjtitle{Nanophotonics}
\bvolume{2}(\bissue{5-6}),
\bfpage{407}--\blpage{428}
(\byear{2013})
\end{barticle}
\endbibitem

\bibitem[\protect\citeauthoryear{Berneschi et~al.}{2020}]{berneschi2020optical}
\begin{barticle}
\bauthor{\bsnm{Berneschi}, \binits{S.}},
\bauthor{\bsnm{Barucci}, \binits{A.}},
\bauthor{\bsnm{Baldini}, \binits{F.}},
\bauthor{\bsnm{Cosi}, \binits{F.}},
\bauthor{\bsnm{Quercioli}, \binits{F.}},
\bauthor{\bsnm{Pelli}, \binits{S.}},
\bauthor{\bsnm{Righini}, \binits{G.C.}},
\bauthor{\bsnm{Tiribilli}, \binits{B.}},
\bauthor{\bsnm{Tombelli}, \binits{S.}},
\bauthor{\bsnm{Trono}, \binits{C.}}, \betal:
\batitle{Optical fibre micro/nano tips as fluorescence-based sensors and
  interrogation probes}.
\bjtitle{Optics}
\bvolume{1}(\bissue{2}),
\bfpage{213}--\blpage{242}
(\byear{2020})
\end{barticle}
\endbibitem

\bibitem[\protect\citeauthoryear{Le~Kien et~al.}{2005}]{le2005spontaneous}
\begin{barticle}
\bauthor{\bsnm{Le~Kien}, \binits{F.}},
\bauthor{\bsnm{Gupta}, \binits{S.D.}},
\bauthor{\bsnm{Balykin}, \binits{V.}},
\bauthor{\bsnm{Hakuta}, \binits{K.}}:
\batitle{Spontaneous emission of a cesium atom near a nanofiber: Efficient
  coupling of light to guided modes}.
\bjtitle{Physical Review A}
\bvolume{72}(\bissue{3}),
\bfpage{032509}
(\byear{2005})
\end{barticle}
\endbibitem

\bibitem[\protect\citeauthoryear{Nayak and Hakuta}{2008}]{nayak2008single}
\begin{barticle}
\bauthor{\bsnm{Nayak}, \binits{K.P.}},
\bauthor{\bsnm{Hakuta}, \binits{K.}}:
\batitle{Single atoms on an optical nanofibre}.
\bjtitle{New Journal of Physics}
\bvolume{10}(\bissue{5}),
\bfpage{053003}
(\byear{2008})
\end{barticle}
\endbibitem

\bibitem[\protect\citeauthoryear{Yalla et~al.}{2012}]{yalla2012efficient}
\begin{barticle}
\bauthor{\bsnm{Yalla}, \binits{R.}},
\bauthor{\bsnm{Le~Kien}, \binits{F.}},
\bauthor{\bsnm{Morinaga}, \binits{M.}},
\bauthor{\bsnm{Hakuta}, \binits{K.}}:
\batitle{Efficient channeling of fluorescence photons from single quantum dots
  into guided modes of optical nanofiber}.
\bjtitle{Physical review letters}
\bvolume{109}(\bissue{6}),
\bfpage{063602}
(\byear{2012})
\end{barticle}
\endbibitem

\bibitem[\protect\citeauthoryear{Chonan et~al.}{2014}]{chonan2014efficient}
\begin{barticle}
\bauthor{\bsnm{Chonan}, \binits{S.}},
\bauthor{\bsnm{Kato}, \binits{S.}},
\bauthor{\bsnm{Aoki}, \binits{T.}}:
\batitle{Efficient single-mode photon-coupling device utilizing a nanofiber
  tip}.
\bjtitle{Scientific reports}
\bvolume{4}(\bissue{1}),
\bfpage{4785}
(\byear{2014})
\end{barticle}
\endbibitem

\bibitem[\protect\citeauthoryear{Resmi et~al.}{2023}]{resmi2023efficient}
\begin{bchapter}
\bauthor{\bsnm{Resmi}, \binits{M.}},
\bauthor{\bsnm{Bashaiah}, \binits{E.}},
\bauthor{\bsnm{Das}, \binits{B.}},
\bauthor{\bsnm{Yalla}, \binits{R.}}:
\bctitle{Efficient fiber-coupled single photon source using an optical
  nanofiber tip}.
In: \bbtitle{Women in Optics and Photonics in India 2022},
vol. \bseriesno{12638},
pp. \bfpage{107}--\blpage{109}
(\byear{2023}).
\bcomment{SPIE}
\end{bchapter}
\endbibitem

\bibitem[\protect\citeauthoryear{Resmi et~al.}{2024}]{resmichanneling}
\begin{barticle}
\bauthor{\bsnm{Resmi}, \binits{M.}},
\bauthor{\bsnm{Bashaiah}, \binits{E.}},
\bauthor{\bsnm{Yalla}, \binits{R.}}:
\batitle{Channeling of fluorescence photons from quantum dots into guided modes
  of an optical nanofiber tip}.
\bjtitle{Journal of Optics}
\bvolume{26}(\bissue{6}),
\bfpage{065401}
(\byear{2024})
\end{barticle}
\endbibitem

\bibitem[\protect\citeauthoryear{Babinec et~al.}{2010}]{babinec2010diamond}
\begin{barticle}
\bauthor{\bsnm{Babinec}, \binits{T.M.}},
\bauthor{\bsnm{Hausmann}, \binits{B.J.}},
\bauthor{\bsnm{Khan}, \binits{M.}},
\bauthor{\bsnm{Zhang}, \binits{Y.}},
\bauthor{\bsnm{Maze}, \binits{J.R.}},
\bauthor{\bsnm{Hemmer}, \binits{P.R.}},
\bauthor{\bsnm{Lon{\v{c}}ar}, \binits{M.}}:
\batitle{A diamond nanowire single-photon source}.
\bjtitle{Nature nanotechnology}
\bvolume{5}(\bissue{3}),
\bfpage{195}--\blpage{199}
(\byear{2010})
\end{barticle}
\endbibitem

\bibitem[\protect\citeauthoryear{Das et~al.}{2023}]{das2023efficient}
\begin{bchapter}
\bauthor{\bsnm{Das}, \binits{B.}},
\bauthor{\bsnm{Resmi}, \binits{M.}},
\bauthor{\bsnm{Bashaiah}, \binits{E.}},
\bauthor{\bsnm{Yalla}, \binits{R.}}:
\bctitle{Efficient single-mode coupling design using a silica/diamond nano-tip
  with a gold nanoparticle}.
In: \bbtitle{Women in Optics and Photonics in India 2022},
vol. \bseriesno{12638},
pp. \bfpage{125}--\blpage{127}
(\byear{2023}).
\bcomment{SPIE}
\end{bchapter}
\endbibitem

\bibitem[\protect\citeauthoryear{Yalla et~al.}{2014}]{yalla2014cavity}
\begin{barticle}
\bauthor{\bsnm{Yalla}, \binits{R.}},
\bauthor{\bsnm{Sadgrove}, \binits{M.}},
\bauthor{\bsnm{Nayak}, \binits{K.P.}},
\bauthor{\bsnm{Hakuta}, \binits{K.}}:
\batitle{Cavity quantum electrodynamics on a nanofiber using a composite
  photonic crystal cavity}.
\bjtitle{Physical review letters}
\bvolume{113}(\bissue{14}),
\bfpage{143601}
(\byear{2014})
\end{barticle}
\endbibitem

\bibitem[\protect\citeauthoryear{Schell et~al.}{2015}]{schell2015highly}
\begin{barticle}
\bauthor{\bsnm{Schell}, \binits{A.W.}},
\bauthor{\bsnm{Takashima}, \binits{H.}},
\bauthor{\bsnm{Kamioka}, \binits{S.}},
\bauthor{\bsnm{Oe}, \binits{Y.}},
\bauthor{\bsnm{Fujiwara}, \binits{M.}},
\bauthor{\bsnm{Benson}, \binits{O.}},
\bauthor{\bsnm{Takeuchi}, \binits{S.}}:
\batitle{Highly efficient coupling of nanolight emitters to a ultra-wide
  tunable nanofibre cavity}.
\bjtitle{Scientific reports}
\bvolume{5}(\bissue{1}),
\bfpage{9619}
(\byear{2015})
\end{barticle}
\endbibitem

\bibitem[\protect\citeauthoryear{Takashima
  et~al.}{2016}]{takashima2016detailed}
\begin{barticle}
\bauthor{\bsnm{Takashima}, \binits{H.}},
\bauthor{\bsnm{Fujiwara}, \binits{M.}},
\bauthor{\bsnm{Schell}, \binits{A.W.}},
\bauthor{\bsnm{Takeuchi}, \binits{S.}}:
\batitle{Detailed numerical analysis of photon emission from a single light
  emitter coupled with a nanofiber bragg cavity}.
\bjtitle{Optics express}
\bvolume{24}(\bissue{13}),
\bfpage{15050}--\blpage{15058}
(\byear{2016})
\end{barticle}
\endbibitem

\bibitem[\protect\citeauthoryear{Patel et~al.}{2016}]{patel2016efficient}
\begin{barticle}
\bauthor{\bsnm{Patel}, \binits{R.N.}},
\bauthor{\bsnm{Schr{\"o}der}, \binits{T.}},
\bauthor{\bsnm{Wan}, \binits{N.}},
\bauthor{\bsnm{Li}, \binits{L.}},
\bauthor{\bsnm{Mouradian}, \binits{S.L.}},
\bauthor{\bsnm{Chen}, \binits{E.H.}},
\bauthor{\bsnm{Englund}, \binits{D.R.}}:
\batitle{Efficient photon coupling from a diamond nitrogen vacancy center by
  integration with silica fiber}.
\bjtitle{Light: Science \& Applications}
\bvolume{5}(\bissue{2}),
\bfpage{16032}--\blpage{16032}
(\byear{2016})
\end{barticle}
\endbibitem

\bibitem[\protect\citeauthoryear{Yonezu et~al.}{2017}]{yonezu2017efficient}
\begin{barticle}
\bauthor{\bsnm{Yonezu}, \binits{Y.}},
\bauthor{\bsnm{Wakui}, \binits{K.}},
\bauthor{\bsnm{Furusawa}, \binits{K.}},
\bauthor{\bsnm{Takeoka}, \binits{M.}},
\bauthor{\bsnm{Semba}, \binits{K.}},
\bauthor{\bsnm{Aoki}, \binits{T.}}:
\batitle{Efficient single-photon coupling from a nitrogen-vacancy center
  embedded in a diamond nanowire utilizing an optical nanofiber}.
\bjtitle{Scientific reports}
\bvolume{7}(\bissue{1}),
\bfpage{12985}
(\byear{2017})
\end{barticle}
\endbibitem

\bibitem[\protect\citeauthoryear{Li et~al.}{2018}]{li2018tailoring}
\begin{barticle}
\bauthor{\bsnm{Li}, \binits{W.}},
\bauthor{\bsnm{Du}, \binits{J.}},
\bauthor{\bsnm{Chormaic}, \binits{S.N.}}:
\batitle{Tailoring a nanofiber for enhanced photon emission and coupling
  efficiency from single quantum emitters}.
\bjtitle{Optics letters}
\bvolume{43}(\bissue{8}),
\bfpage{1674}--\blpage{1677}
(\byear{2018})
\end{barticle}
\endbibitem

\bibitem[\protect\citeauthoryear{Nayak et~al.}{2019}]{nayak2019real}
\begin{barticle}
\bauthor{\bsnm{Nayak}, \binits{K.P.}},
\bauthor{\bsnm{Wang}, \binits{J.}},
\bauthor{\bsnm{Keloth}, \binits{J.}}:
\batitle{Real-time observation of single atoms trapped and interfaced to a
  nanofiber cavity}.
\bjtitle{Physical Review Letters}
\bvolume{123}(\bissue{21}),
\bfpage{213602}
(\byear{2019})
\end{barticle}
\endbibitem

\bibitem[\protect\citeauthoryear{Wang et~al.}{2021}]{wang2021high}
\begin{barticle}
\bauthor{\bsnm{Wang}, \binits{X.}},
\bauthor{\bsnm{Zhang}, \binits{P.}},
\bauthor{\bsnm{Li}, \binits{G.}},
\bauthor{\bsnm{Zhang}, \binits{T.}}:
\batitle{High-efficiency coupling of single quantum emitters into hole-tailored
  nanofibers}.
\bjtitle{Optics Express}
\bvolume{29}(\bissue{7}),
\bfpage{11158}--\blpage{11168}
(\byear{2021})
\end{barticle}
\endbibitem

\bibitem[\protect\citeauthoryear{Yalla et~al.}{2022}]{yalla2022integration}
\begin{botherref}
\oauthor{\bsnm{Yalla}, \binits{R.}},
\oauthor{\bsnm{Kojima}, \binits{Y.}},
\oauthor{\bsnm{Fukumoto}, \binits{Y.}},
\oauthor{\bsnm{Suzuki}, \binits{H.}},
\oauthor{\bsnm{Ariyada}, \binits{O.}},
\oauthor{\bsnm{Shafi}, \binits{K.M.}},
\oauthor{\bsnm{Nayak}, \binits{K.P.}},
\oauthor{\bsnm{Hakuta}, \binits{K.}}:
Integration of silicon-vacancy centers in nanodiamonds with an optical
  nanofiber.
Applied Physics Letters
\textbf{120}(24)
(2022)
\end{botherref}
\endbibitem

\bibitem[\protect\citeauthoryear{Sipahigil
  et~al.}{2016}]{sipahigil2016integrated}
\begin{barticle}
\bauthor{\bsnm{Sipahigil}, \binits{A.}},
\bauthor{\bsnm{Evans}, \binits{R.E.}},
\bauthor{\bsnm{Sukachev}, \binits{D.D.}},
\bauthor{\bsnm{Burek}, \binits{M.J.}},
\bauthor{\bsnm{Borregaard}, \binits{J.}},
\bauthor{\bsnm{Bhaskar}, \binits{M.K.}},
\bauthor{\bsnm{Nguyen}, \binits{C.T.}},
\bauthor{\bsnm{Pacheco}, \binits{J.L.}},
\bauthor{\bsnm{Atikian}, \binits{H.A.}},
\bauthor{\bsnm{Meuwly}, \binits{C.}}, \betal:
\batitle{An integrated diamond nanophotonics platform for quantum-optical
  networks}.
\bjtitle{Science}
\bvolume{354}(\bissue{6314}),
\bfpage{847}--\blpage{850}
(\byear{2016})
\end{barticle}
\endbibitem

\bibitem[\protect\citeauthoryear{Nguyen et~al.}{2019}]{nguyen2019quantum}
\begin{barticle}
\bauthor{\bsnm{Nguyen}, \binits{C.}},
\bauthor{\bsnm{Sukachev}, \binits{D.}},
\bauthor{\bsnm{Bhaskar}, \binits{M.}},
\bauthor{\bsnm{Machielse}, \binits{B.}},
\bauthor{\bsnm{Levonian}, \binits{D.}},
\bauthor{\bsnm{Knall}, \binits{E.}},
\bauthor{\bsnm{Stroganov}, \binits{P.}},
\bauthor{\bsnm{Riedinger}, \binits{R.}},
\bauthor{\bsnm{Park}, \binits{H.}},
\bauthor{\bsnm{Lon{\v{c}}ar}, \binits{M.}}, \betal:
\batitle{Quantum network nodes based on diamond qubits with an efficient
  nanophotonic interface}.
\bjtitle{Physical review letters}
\bvolume{123}(\bissue{18}),
\bfpage{183602}
(\byear{2019})
\end{barticle}
\endbibitem

\bibitem[\protect\citeauthoryear{Schr{\"o}der
  et~al.}{2016}]{schroder2016quantum}
\begin{barticle}
\bauthor{\bsnm{Schr{\"o}der}, \binits{T.}},
\bauthor{\bsnm{Mouradian}, \binits{S.L.}},
\bauthor{\bsnm{Zheng}, \binits{J.}},
\bauthor{\bsnm{Trusheim}, \binits{M.E.}},
\bauthor{\bsnm{Walsh}, \binits{M.}},
\bauthor{\bsnm{Chen}, \binits{E.H.}},
\bauthor{\bsnm{Li}, \binits{L.}},
\bauthor{\bsnm{Bayn}, \binits{I.}},
\bauthor{\bsnm{Englund}, \binits{D.}}:
\batitle{Quantum nanophotonics in diamond}.
\bjtitle{JOSA B}
\bvolume{33}(\bissue{4}),
\bfpage{65}--\blpage{83}
(\byear{2016})
\end{barticle}
\endbibitem

\bibitem[\protect\citeauthoryear{Ruf et~al.}{2021}]{ruf2021quantum}
\begin{botherref}
\oauthor{\bsnm{Ruf}, \binits{M.}},
\oauthor{\bsnm{Wan}, \binits{N.H.}},
\oauthor{\bsnm{Choi}, \binits{H.}},
\oauthor{\bsnm{Englund}, \binits{D.}},
\oauthor{\bsnm{Hanson}, \binits{R.}}:
Quantum networks based on color centers in diamond.
Journal of Applied Physics
\textbf{130}(7)
(2021)
\end{botherref}
\endbibitem

\bibitem[\protect\citeauthoryear{Burek et~al.}{2017}]{burek2017fiber}
\begin{barticle}
\bauthor{\bsnm{Burek}, \binits{M.J.}},
\bauthor{\bsnm{Meuwly}, \binits{C.}},
\bauthor{\bsnm{Evans}, \binits{R.E.}},
\bauthor{\bsnm{Bhaskar}, \binits{M.K.}},
\bauthor{\bsnm{Sipahigil}, \binits{A.}},
\bauthor{\bsnm{Meesala}, \binits{S.}},
\bauthor{\bsnm{Machielse}, \binits{B.}},
\bauthor{\bsnm{Sukachev}, \binits{D.D.}},
\bauthor{\bsnm{Nguyen}, \binits{C.T.}},
\bauthor{\bsnm{Pacheco}, \binits{J.L.}}, \betal:
\batitle{Fiber-coupled diamond quantum nanophotonic interface}.
\bjtitle{Physical Review Applied}
\bvolume{8}(\bissue{2}),
\bfpage{024026}
(\byear{2017})
\end{barticle}
\endbibitem

\bibitem[\protect\citeauthoryear{Siampour et~al.}{2020}]{Siampour}
\begin{barticle}
\bauthor{\bsnm{Siampour}, \binits{H.}},
\bauthor{\bsnm{Wang}, \binits{O.}},
\bauthor{\bsnm{Zenin}, \binits{V.A.}},
\bauthor{\bsnm{Boroviks}, \binits{S.}},
\bauthor{\bsnm{Siyushev}, \binits{P.}},
\bauthor{\bsnm{Yang}, \binits{Y.}},
\bauthor{\bsnm{Davydov}, \binits{V.A.}},
\bauthor{\bsnm{Kulikova}, \binits{L.F.}},
\bauthor{\bsnm{Agafonov}, \binits{V.N.}},
\bauthor{\bsnm{Kubanek}, \binits{A.}},
\bauthor{\bsnm{Mortensen}, \binits{N.A.}},
\bauthor{\bsnm{Jelezko}, \binits{F.}},
\bauthor{\bsnm{Bozhevolnyi}, \binits{S.I.}}:
\batitle{Ultrabright single-photon emission from germanium-vacancy zero-phonon
  lines: deterministic emitter-waveguide interfacing at plasmonic hot spots}.
\bjtitle{Nanophotonics}
\bvolume{9}(\bissue{4}),
\bfpage{953}--\blpage{962}
(\byear{2020})
\end{barticle}
\endbibitem

\bibitem[\protect\citeauthoryear{Wang et~al.}{2014}]{wang2014gallium}
\begin{barticle}
\bauthor{\bsnm{Wang}, \binits{J.}},
\bauthor{\bsnm{Santamato}, \binits{A.}},
\bauthor{\bsnm{Jiang}, \binits{P.}},
\bauthor{\bsnm{Bonneau}, \binits{D.}},
\bauthor{\bsnm{Engin}, \binits{E.}},
\bauthor{\bsnm{Silverstone}, \binits{J.W.}},
\bauthor{\bsnm{Lermer}, \binits{M.}},
\bauthor{\bsnm{Beetz}, \binits{J.}},
\bauthor{\bsnm{Kamp}, \binits{M.}},
\bauthor{\bsnm{H{\"o}fling}, \binits{S.}}, \betal:
\batitle{Gallium arsenide (gaas) quantum photonic waveguide circuits}.
\bjtitle{Optics Communications}
\bvolume{327},
\bfpage{49}--\blpage{55}
(\byear{2014})
\end{barticle}
\endbibitem

\bibitem[\protect\citeauthoryear{Reimer et~al.}{2012}]{reimer2012bright}
\begin{barticle}
\bauthor{\bsnm{Reimer}, \binits{M.E.}},
\bauthor{\bsnm{Bulgarini}, \binits{G.}},
\bauthor{\bsnm{Akopian}, \binits{N.}},
\bauthor{\bsnm{Hocevar}, \binits{M.}},
\bauthor{\bsnm{Bavinck}, \binits{M.B.}},
\bauthor{\bsnm{Verheijen}, \binits{M.A.}},
\bauthor{\bsnm{Bakkers}, \binits{E.P.}},
\bauthor{\bsnm{Kouwenhoven}, \binits{L.P.}},
\bauthor{\bsnm{Zwiller}, \binits{V.}}:
\batitle{Bright single-photon sources in bottom-up tailored nanowires}.
\bjtitle{Nature communications}
\bvolume{3}(\bissue{1}),
\bfpage{737}
(\byear{2012})
\end{barticle}
\endbibitem

\bibitem[\protect\citeauthoryear{Le~Kien et~al.}{2020}]{le2020coupling}
\begin{barticle}
\bauthor{\bsnm{Le~Kien}, \binits{F.}},
\bauthor{\bsnm{Ruks}, \binits{L.}},
\bauthor{\bsnm{Chormaic}, \binits{S.N.}},
\bauthor{\bsnm{Busch}, \binits{T.}}:
\batitle{Coupling between guided modes of two parallel nanofibers}.
\bjtitle{New Journal of Physics}
\bvolume{22}(\bissue{12}),
\bfpage{123007}
(\byear{2020})
\end{barticle}
\endbibitem

\bibitem[\protect\citeauthoryear{Le~Kien et~al.}{2021}]{le2021spatial}
\begin{barticle}
\bauthor{\bsnm{Le~Kien}, \binits{F.}},
\bauthor{\bsnm{Ruks}, \binits{L.}},
\bauthor{\bsnm{Chormaic}, \binits{S.N.}},
\bauthor{\bsnm{Busch}, \binits{T.}}:
\batitle{Spatial distributions of the fields in guided normal modes of two
  coupled parallel optical nanofibers}.
\bjtitle{New Journal of Physics}
\bvolume{23}(\bissue{4}),
\bfpage{043006}
(\byear{2021})
\end{barticle}
\endbibitem

\bibitem[\protect\citeauthoryear{Shao et~al.}{2022}]{shao2022twin}
\begin{barticle}
\bauthor{\bsnm{Shao}, \binits{L.}},
\bauthor{\bsnm{Wu}, \binits{H.}},
\bauthor{\bsnm{Fang}, \binits{W.}},
\bauthor{\bsnm{Tong}, \binits{L.}}:
\batitle{Twin-nanofiber structure for a highly efficient single-photon
  collection}.
\bjtitle{Optics Express}
\bvolume{30}(\bissue{6}),
\bfpage{9147}--\blpage{9155}
(\byear{2022})
\end{barticle}
\endbibitem

\bibitem[\protect\citeauthoryear{Yang et~al.}{2024}]{yang2024generating}
\begin{barticle}
\bauthor{\bsnm{Yang}, \binits{Y.}},
\bauthor{\bsnm{Gao}, \binits{J.}},
\bauthor{\bsnm{Wu}, \binits{H.}},
\bauthor{\bsnm{Zhou}, \binits{Z.}},
\bauthor{\bsnm{Yang}, \binits{L.}},
\bauthor{\bsnm{Guo}, \binits{X.}},
\bauthor{\bsnm{Wang}, \binits{P.}},
\bauthor{\bsnm{Tong}, \binits{L.}}:
\batitle{Generating a nanoscale blade-like optical field in a coupled nanofiber
  pair}.
\bjtitle{Photonics Research}
\bvolume{12}(\bissue{1}),
\bfpage{154}--\blpage{162}
(\byear{2024})
\end{barticle}
\endbibitem

\bibitem[\protect\citeauthoryear{Garcia et~al.}{2005}]{garcia2005new}
\begin{barticle}
\bauthor{\bsnm{Garcia}, \binits{S.G.}},
\bauthor{\bsnm{Bretones}, \binits{A.R.}},
\bauthor{\bsnm{Olmedo}, \binits{B.G.}},
\bauthor{\bsnm{Martin}, \binits{R.G.}}:
\batitle{New trends in fdtd methods in computational electrodynamics:
  Unconditionally stable schemes}.
\bjtitle{Recent Res. Development in Electronics}
\bvolume{2},
\bfpage{55}--\blpage{96}
(\byear{2005})
\end{barticle}
\endbibitem

\bibitem[\protect\citeauthoryear{Resmi et~al.}{2024}]{resmifabrication}
\begin{barticle}
\bauthor{\bsnm{Resmi}, \binits{M.}},
\bauthor{\bsnm{Elaganuru}, \binits{B.}},
\bauthor{\bsnm{Ramachandrarao}, \binits{Y.}}:
\batitle{Fabrication and characterization of optical nanofiber tips}.
\bjtitle{Journal of Nanophotonics}
\bvolume{18}(\bissue{2}),
\bfpage{026007}
(\byear{2024})
\end{barticle}
\endbibitem

\bibitem[\protect\citeauthoryear{Burek et~al.}{2012}]{burek2012free}
\begin{barticle}
\bauthor{\bsnm{Burek}, \binits{M.J.}},
\bauthor{\bsnm{De~Leon}, \binits{N.P.}},
\bauthor{\bsnm{Shields}, \binits{B.J.}},
\bauthor{\bsnm{Hausmann}, \binits{B.J.}},
\bauthor{\bsnm{Chu}, \binits{Y.}},
\bauthor{\bsnm{Quan}, \binits{Q.}},
\bauthor{\bsnm{Zibrov}, \binits{A.S.}},
\bauthor{\bsnm{Park}, \binits{H.}},
\bauthor{\bsnm{Lukin}, \binits{M.D.}},
\bauthor{\bsnm{Loncar}, \binits{M.}}:
\batitle{Free-standing mechanical and photonic nanostructures in single-crystal
  diamond}.
\bjtitle{Nano letters}
\bvolume{12}(\bissue{12}),
\bfpage{6084}--\blpage{6089}
(\byear{2012})
\end{barticle}
\endbibitem

\bibitem[\protect\citeauthoryear{Burek et~al.}{2016}]{burek2016diamond}
\begin{barticle}
\bauthor{\bsnm{Burek}, \binits{M.J.}},
\bauthor{\bsnm{Cohen}, \binits{J.D.}},
\bauthor{\bsnm{Meenehan}, \binits{S.M.}},
\bauthor{\bsnm{El-Sawah}, \binits{N.}},
\bauthor{\bsnm{Chia}, \binits{C.}},
\bauthor{\bsnm{Ruelle}, \binits{T.}},
\bauthor{\bsnm{Meesala}, \binits{S.}},
\bauthor{\bsnm{Rochman}, \binits{J.}},
\bauthor{\bsnm{Atikian}, \binits{H.A.}},
\bauthor{\bsnm{Markham}, \binits{M.}}, \betal:
\batitle{Diamond optomechanical crystals}.
\bjtitle{Optica}
\bvolume{3}(\bissue{12}),
\bfpage{1404}--\blpage{1411}
(\byear{2016})
\end{barticle}
\endbibitem

\end{thebibliography}

\end{document}